\documentclass[11pt]{article}
\usepackage{authblk}
\usepackage{graphicx}
\usepackage{epstopdf}
\usepackage{latexsym}
\usepackage{multirow}
\usepackage{tabls}
\usepackage{epstopdf}
\usepackage{graphicx}
\usepackage{amsmath, amsthm, amssymb}
\begin{document}

\title{A General Slice Moment Decomposition of RMS Beam Emittance}
\author{Chad Mitchell}
\affil{{Lawrence Berkeley National Laboratory, Berkeley, California, 94720, USA}}

\maketitle

\begin{abstract}
The square of the horizontal projected (rms) beam emittance is expressed as the sum of four nonnegative contributions, each described using the slice moments of the beam and possessing a natural interpretation in terms of the geometrical properties of the beam in the six-dimensional phase space.  The mathematical formalism describing the relationships between projected beam quantities and slice beam quantities is reviewed.  The results may be used to reconstruct the moments and emittances of the beam from the moments of its subpopulations, as well as to isolate and better understand a variety of slice and interslice dynamical contributions to the projected beam emittance growth.
\end{abstract}

\section{Introduction}
The concepts of slice and projected beam emittance play a central role in the design and optimization of next-generation FEL light sources \cite{LightSource}, in the theory of emittance compensation \cite{EmittanceComp1}-\cite{EmittanceComp2}, and in the general theory of beam dynamics in linear and circular accelerators \cite{Reiser}.  The slice formalism is particularly useful in the case of long beams with small uncorrelated energy spread, for which the longitudinal slice-to-slice variation in the fields experienced by particles in the beam (e.g., due to wakefields and time-dependent RF fields) can be significant, while there is little particle slippage between slices.  In this case, the dynamics of particles within each slice may profitably be separated from the dynamics of the slice centroids.  

This separation leads naturally to the question of the relationship between the projected (rms) emittance and the properties of the various beam slices.  We demonstrate that the square of the projected emittance can be decomposed as the sum of four contributions, each with a distinct geometrical interpretation in terms of the slice beam moments.  Fragments of a treatment along these lines can be found in various forms in the literature \cite{Beamlet}-\cite{Mitchell}.  The purpose of this paper is to consolidate these results in the form of a general and carefully presented mathematical treatment, with proofs of some nontrivial features included.  Several tools are presented that point the way toward a similar treatment of other diagnostic beam quantities such as the projected beam energy spread, the 4D and 6D emittances, and the eigenemittances for coupled beams \cite{Eigenemittance}.

\section{Basic formalism}
Let $f$ denote a beam distribution function on the 6D phase space, which we parameterize by canonical coordinates ${\bf X}=(x,p_x,y,p_y,z,p_z)$.  For convenience, we will assume that all momenta are normalized by the quantity $mc$.  By assumption, $f$ has the properties that $f\geq 0$ and
\begin{equation}
\int f({\bf X})d{\bf X}=1.
\end{equation}
Thus, $f$ is a joint probability density in the six random variables $x$, $p_x$, $y$, $p_y$, $z$, and $p_z$ that characterizes the distribution of the beam particles.  If $\phi$ is any function defined on the phase space, we define its beam average in the natural way:
\begin{equation}
\langle{\phi\rangle}=\int \phi({\bf X})f({\bf X})d{\bf X}. \label{avg}
\end{equation}
Assume that $z$ is the direction of beam motion, and consider the planes of constant $z$, which we describe as {\it longitudinal slices}.  Let ${\bf \zeta}=(x,p_x,y,p_y,p_z)$ denote the remaining phase space coordinates, so we will write ${\bf X}=(z,\zeta)$ when we wish to emphasize the separation of the slice coordinate $z$ from the coordinates within a slice $\zeta$.  There are two natural functions associated with this decomposition into slices.  
First, the {\it longitudinal density profile} of the beam is given by the following marginal probability density:
\begin{equation}
\lambda(z)=\int f(z,\zeta)d\zeta. \label{longdens}
\end{equation}
Second, the {\it slice distribution function} for the slice $z$ is given by the conditional probability density:
\begin{equation}
f({\bf \zeta}|z)=
\frac{f(z,{\bf \zeta})}{\int f(z,{\bf \zeta})d{\bf\zeta}}=\frac{f(z,{\bf\zeta})}{\lambda(z)}, \quad\text{provided}\quad \lambda(z)>0, \label{slicedist}
\end{equation}
with $f(\zeta |z)=0$ when $\lambda(z)=0$.  It follows that both of these quantities are nonnegative and normalized so that:
\begin{equation}
\int\lambda(z)dz=1,\quad\quad \int f(\zeta|z)d\zeta=1.
\end{equation}
Given any function $\phi$ defined on the phase space, we may define its average within slice $z$ by using the slice distribution function:
\begin{equation}
\langle{\phi\rangle}_z=\int\phi(z,\zeta)f({\bf \zeta}|z)d\zeta. \label{sliceavg}
\end{equation}
We use the notation $f|z$ to refer to the the slice distribution function when we wish to avoid listing its arguments, in order to distinguish this quantity from the full beam distribution function $f$.

\section{Slice and projected moments}
We use the notation $\vec{\mu}$ and $\Sigma$ to denote the centroid vector and covariance matrix of the beam, respectively, whose elements are given by:
\begin{equation}
\mu_a=\langle{X_a\rangle},\quad\quad \Sigma_{ab}=\langle{(X_a-\mu_a)(X_b-\mu_b)\rangle}\quad\quad \text{for}\quad a,b=1,\ldots,6. \label{projmoment}
\end{equation}
Likewise, define the slice centroid vector $\vec{\mu}(z)$ and slice covariance matrix $\Sigma(z)$ for slice $z$ by:
\begin{equation}
\mu_a(z)=\langle{X_a\rangle}_z,\quad\quad \Sigma_{ab}(z)=\langle{(X_a-\mu_a(z))(X_b-\mu_b(z))\rangle}_z \quad\quad \text{for}\quad a=1,\ldots,6. \label{slicemoment}
\end{equation}
Note that we have used two distinct forms of averaging $\langle{\rangle}$ and $\langle{\rangle}_z$, associated with the probability densities $f$ and $f|z$.  
Given a quantity $\phi$ that depends on $z$, we may also define 
 an average across slices by using the longitudinal density profile $\lambda$:
\begin{equation}
\operatorname{E}[\phi]=\int \phi(z)\lambda(z)dz.
\end{equation}
Likewise, given two $z-$dependent quantities $\phi_1$ and $\phi_2$, we define their covariance in analogy with (\ref{projmoment}) and (\ref{slicemoment}):
\begin{equation}
\operatorname{Cov}[\phi_1,\phi_2]=\operatorname{E}[(\phi_1-\operatorname{E}[\phi_1])(\phi_2-\operatorname{E}[\phi_2])], \label{zCovar}
\end{equation}
and the variance is defined in the usual way as:
\begin{equation}
\operatorname{Var}[\phi]=\operatorname{Cov}[\phi,\phi]=\operatorname{E}[(\phi-\operatorname{E}[\phi])^2].
\end{equation}
The projected (\ref{projmoment}) and slice (\ref{slicemoment}) moments of the beam can now be related using two fundamental results from the theory of conditional probability.  The {\it law of total expectation} gives:
\begin{equation}
\mu_a=\operatorname{E}[\mu_a(z)]\quad\quad \text{for}\quad a=1,\ldots,6, \label{expect}
\end{equation}
and the {\it law of total covariance} gives:
\begin{equation}
\Sigma_{ab}=\operatorname{E}[\Sigma_{ab}(z)]+\operatorname{Cov}[\mu_a(z),\mu_b(z)]\quad\quad \text{for}\quad a,b=1,\ldots,6. \label{covar}
\end{equation}
Note that the notation in (\ref{expect}-\ref{covar}) makes sense, because the slice moments appearing within the brackets $[\cdot ]$ on each right-hand side are $z-$dependent quantities.  A proof of each of these claims is provided in Appendix A.  

The projected (rms) normalized emittance of the beam in the $(x,p_x)$ plane is typically defined by:
\begin{equation}
\epsilon_x^2=\langle{(x-\langle{x\rangle})^2\rangle}\langle{(p_x-\langle{p_x\rangle})^2\rangle}-\langle{(x-\langle{x\rangle})(p_x-\langle{p_x\rangle})\rangle}^2, \label{projemit}
\end{equation}
with expressions of the same form for $\epsilon_y^2$ and $\epsilon_z^2$.  It is convenient to express these definitions using the 2$\times$2 blocks of the 6D covariance matrix associated with each degree of freedom:
\begin{equation}
\Sigma^X=\begin{pmatrix}
\Sigma_{11} & \Sigma_{12} \\
\Sigma_{21} & \Sigma_{22}
\end{pmatrix}\quad\quad
\Sigma^Y=\begin{pmatrix}
\Sigma_{33} & \Sigma_{34} \\
\Sigma_{43} & \Sigma_{44}
\end{pmatrix}\quad\quad
\Sigma^Z=\begin{pmatrix}
\Sigma_{55} & \Sigma_{56} \\
\Sigma_{65} & \Sigma_{66}
\end{pmatrix},
\end{equation}
so that:
\begin{equation}
\epsilon_x^2=\det\Sigma^X,\quad \epsilon_y^2=\det\Sigma^Y,\quad \epsilon_z^2=\det{\Sigma^Z}. \label{detdef}
\end{equation}
In the same way, we may express the emittances within slice $z$ as:
\begin{equation}
\epsilon_x^2(z)=\det\Sigma^X(z),\quad \epsilon_y^2=\det\Sigma^Y(z),\quad \epsilon_z^2=\det{\Sigma^Z(z)},
\end{equation}
where $\Sigma^X(z)$, $\Sigma^Y(z)$, and $\Sigma^Z(z)$ are obtained from the slice covariance matrix $\Sigma(z)$ defined in (\ref{slicemoment}).  It follows from our definitions that $\epsilon_z(z)=0$ for all $z$, so this quantity will play no role in the discussions to follow.

\section{Decomposition of the projected emittance}
In this section we will show that the horizontal projected emittance (\ref{projemit}) can be written as the sum in quadrature of four contributions, each with a natural interpretation in terms of the slice properties of the beam.  Corresponding results apply also for the projected emittances in the other two planes. 

Define a 6$\times$6 matrix $\Lambda$ according to:
\begin{equation}
\Lambda_{ab}=\operatorname{Cov}[\mu_a(z),\mu_b(z)],\quad\quad \text{for}\quad a,b=1,\ldots,6 \label{lambda}
\end{equation}
and likewise define the 2$\times$2 blocks of $\Lambda$ associated with each degree of freedom:
\begin{equation}
\Lambda^X=\begin{pmatrix}
\Lambda_{11} & \Lambda_{12} \\
\Lambda_{21} & \Lambda_{22}
\end{pmatrix}\quad\quad
\Lambda^Y=\begin{pmatrix}
\Lambda_{33} & \Lambda_{34} \\
\Lambda_{43} & \Lambda_{44}
\end{pmatrix}\quad\quad
\Lambda^Z=\begin{pmatrix}
\Lambda_{55} & \Lambda_{56} \\
\Lambda_{65} & \Lambda_{66}
\end{pmatrix}.
\end{equation}
Then it follows from (\ref{covar}) that we have the 2$\times$2 matrix identities:
\begin{equation}
\Sigma^X=\operatorname{E}[\Sigma^X(z)]+\Lambda^X,\quad \Sigma^Y=\operatorname{E}[\Sigma^Y(z)]+\Lambda^Y,\quad \Sigma^Z=\operatorname{E}[\Sigma^Z(z)]+\Lambda^Z.
\end{equation}
Therefore according to (\ref{detdef}) the horizontal projected emittance is given by:
\begin{equation}
\epsilon_x^2=\operatorname{det}\left(\operatorname{E}[\Sigma^X(z)]+\Lambda^X\right).
\end{equation}
We exploit the following useful identity that applies for any sum of 2$\times$2 matrices, which the reader may directly verify:
\begin{equation}
\operatorname{det}(A+B)=\operatorname{det}(A)+\operatorname{det}(B)+\operatorname{tr}(A)\operatorname{tr}(B)-\operatorname{tr}(AB).
\end{equation}
This gives:
\begin{align}
\epsilon_x^2&=\operatorname{det}(\operatorname{E}[\Sigma^X(z)])+\operatorname{det}(\Lambda^X)+\operatorname{tr}(\operatorname{E}[\Sigma^X(z)])\operatorname{tr}(\Lambda^X)-\operatorname{tr}(\operatorname{E}[\Sigma^X(z)]\Lambda^X) \notag \\
&=\operatorname{det}(\operatorname{E}[\Sigma^X(z)])+\epsilon_{\rm int}^2+\epsilon_{||}^2, \label{step1}
\end{align}
where we have defined:
\begin{align}
\epsilon_{||}^2&=\operatorname{det}\Lambda^X, \label{par} \\
\epsilon_{\rm int}^2&=\operatorname{tr}(\operatorname{E}[\Sigma^X(z)])\operatorname{tr}(\Lambda^X)-\operatorname{tr}(\operatorname{E}[\Sigma^X(z)]\Lambda^X). \label{int}
\end{align}
The significance of these quantities will soon become apparent.
We further decompose the first term appearing on the right-hand side of (\ref{step1}) as follows:
\begin{align}
\operatorname{det}(\operatorname{E}[\Sigma^X(z)])&=\operatorname{E}[\Sigma_{11}(z)]\operatorname{E}[\Sigma_{22}(z)]-\operatorname{E}[\Sigma_{12}(z)]^2 \notag \\
&=\operatorname{E}[\Sigma_{11}(z)\Sigma_{22}(z)]-\operatorname{Cov}[\Sigma_{11}(z),\Sigma_{22}(z)]+\operatorname{Var}[\Sigma_{12}(z)]-\operatorname{E}[\Sigma_{12}(z)^2] \notag \\
&=\operatorname{E}[\epsilon_x^2(z)]-\operatorname{Cov}[\Sigma_{11}(z),\Sigma_{22}(z)]+\operatorname{Var}[\Sigma_{12}(z)] \notag \\
&=\operatorname{E}[\epsilon_x(z)]^2+\operatorname{Var}[\epsilon_x(z)]-\operatorname{Cov}[\Sigma_{11}(z),\Sigma_{22}(z)]+\operatorname{Var}[\Sigma_{12}(z)] \notag \\
&=\epsilon_{\perp}^2+\epsilon_R^2, \label{eperp}
\end{align}
where:
\begin{align}
\epsilon_{\perp}^2&=\operatorname{E}[\epsilon_x(z)]^2, \\
\epsilon_R^2&=\operatorname{Var}[\epsilon_x(z)]-\operatorname{Cov}[\Sigma_{11}(z),\Sigma_{22}(z)]+\operatorname{Var}[\Sigma_{12}(z)].
\end{align}
Thus, we have shown that:
\begin{equation}
\epsilon_x^2=\epsilon_{\perp}^2+\epsilon_R^2+\epsilon_{\rm int}^2+\epsilon_{||}^2.
\end{equation}
In Appendix B, we prove that each of these four contributions is nonnegative, so that $\epsilon_{\perp}$, $\epsilon_R$, $\epsilon_{\rm int}$ and $\epsilon_{||}$ are always real quantities.  

Changing to a more intuitive notation, let us write the first moments of slice $z$ in the form:
\begin{equation}
\mu_x(z)=\mu_1(z) \quad\quad \mu_{p_x}(z)=\mu_2(z),
\end{equation}
which denote the horizontal position and momentum of the slice centroid.  Similarly, let us write the second moments of the slice $z$ in the form:
\begin{equation}
\sigma_x^2(z)=\Sigma_{11}(z), \quad\quad \sigma_{p_x}^2(z)=\Sigma_{22}(z), \quad\quad \langle{\Delta x\Delta p_x\rangle}_z=\Sigma_{12}(z).
\end{equation}
Expressing the four emittance contributions in terms of these quantities gives, after writing out (\ref{par}-\ref{int}) explicitly:
\begin{subequations}\label{decomp}
\begin{align}
\epsilon_{\perp}^2&=\operatorname{E}[\epsilon_x(z)]^2, \\
\epsilon_R^2&=\operatorname{Var}[\epsilon_x(z)]-\operatorname{Cov}[\sigma_x^2(z),\sigma_{p_x}^2(z)]+\operatorname{Var}[\langle{\Delta x\Delta p_x\rangle}_z], \\
\epsilon_{\rm int}^2&=\operatorname{E}[\sigma_x^2(z)]\operatorname{Var}[\mu_{p_x}(z)]+\operatorname{E}[\sigma_{p_x}^2(z)]\operatorname{Var}[\mu_x(z)]-2\operatorname{E}[\langle{xp_x\rangle}_z]
\operatorname{Cov}[\mu_x(z),\mu_{p_x}(z)], \\
\epsilon_{||}^2&=\operatorname{Var}[\mu_x(z)]\operatorname{Var}[\mu_{p_x}(z)]-\operatorname{Cov}[\mu_x(z),\mu_{p_x}(z)]^2.
\end{align}
\end{subequations}
Let us consider each of these contributions in turn.

1. The quantity
\begin{equation}
\epsilon_{\perp}=\operatorname{E}[\epsilon_x(z)] \notag
\end{equation}
is the {\it mean slice emittance}.  It is determined by computing the emittance within each longitudinal beam slice, and then taking the mean among slices, with each slice weighted by its corresponding longitudinal density.  It vanishes if and only if every slice of the beam has zero emittance.

2.  The quantity $\epsilon_R$ is the {\it mismatch emittance}, and its value is determined by the size of variation in the beam second moments (the covariance matrices $\Sigma^X(z)$) from slice to slice.  
When $\epsilon_x(z)\neq 0$, it is typical to define the slice Twiss functions $\alpha(z)$, $\beta(z)$, and $\gamma(z)$ such that\footnote{Each mention of the Twiss functions in this work refers to the horizontal Twiss functions, so the subscripts in $\alpha_x$, $\beta_x$, and $\gamma_x$ have been suppressed; there is no possiblity of confusion with the relativistic factors ${\beta}$ and ${\gamma}$, since these appear only once in (\ref{relativistic}).  Since the emittance in (\ref{Twiss1}) is normalized, the Twiss functions here differ from the geometric Twiss functions by the appropriate relativistic factor.}:
\begin{equation}
\Sigma^X(z)=\epsilon_x(z)
\begin{pmatrix}
\beta(z) & -\alpha(z) \\
-\alpha(z) & \gamma(z)
\end{pmatrix}\quad\text{with}\quad \beta(z)\gamma(z)-\alpha(z)^2=1. \label{Twiss1}
\end{equation}
If the slice Twiss functions $\alpha(z)$, $\beta(z)$, and $\gamma(z)$ are identical for every slice $z$, then $\epsilon_R$ vanishes.

3.  The quantity $\epsilon_{\rm int}$ is the {\it linear misalignment emittance}, and its value is determined by the size of variation in the first moments (slice centroids) from slice to slice.  This is a ``cross term", as the contributions from the centroid variations are themselves weighted by the mean second moments.  If the centroids of all beam slices are aligned, so that $\mu_x(z)$ and $\mu_{p_x}(z)$ are identical for every slice, then $\epsilon_{\rm int}$ vanishes.  

4. The quantity $\epsilon_{||}$ is the {\it nonlinear misalignment emittance}, and its value is also determined by the size of variation in the first moments (slice centroids) from slice to slice.  Unlike $\epsilon_{\rm int}$, it depends exclusively on the slice centroid coordinates.  In particular, $\epsilon_{||}$ is the beam emittance that would result if all charge within each slice were concentrated at the location of the corresponding slice centroid.  If all slice centroids are aligned, then $\epsilon_{||}$ vanishes.  

Appendix B describes the conditions under which each of these contributions vanishes.  The reason for separating the terms $\epsilon_{\rm int}$ and $\epsilon_{||}$ will become more clear in the following section.

\section{Special cases}
The decomposition of the previous section is valid for a general beam distribution.  However, there are several idealized special cases that arise frequently in the literature.  We will consider four of these in this section, with the hope that this will clarify the significance of the four contributions given in (\ref{decomp}).  


\subsection{Uniform slice emittance}
First, suppose that the emittance of every longitudinal beam slice is identical.  More precisely, assume that:
\begin{equation}
\epsilon_x(z)=\operatorname{E}[\epsilon_x(z)]=\epsilon_{\perp} \label{mean}
\end{equation}
for every $z$ with $\lambda(z)\neq 0$.  Provided $\epsilon_{\perp}\neq 0$, it is natural to define the slice Twiss functions $\alpha(z)$, $\beta(z)$, and $\gamma(z)$ as in (\ref{Twiss1}), 
so that
\begin{equation}
\sigma_x^2(z)=\epsilon_{\perp}\beta(z),\quad \sigma_{p_x}^2(z)=\epsilon_{\perp}\gamma(z),\quad \langle{\Delta x\Delta p_x\rangle}_z=-\epsilon_{\perp}\alpha(z). \label{Twiss2}
\end{equation}
In this case the contribution $\epsilon_{\perp}$ is given simply by (\ref{mean}), and the contribution $\epsilon_{||}$ is unchanged from its original form (\ref{decomp}).  We can re-express the remaining two contributions in terms of the Twiss functions as:
\begin{subequations}
\begin{align}
\epsilon_R^2&=\epsilon_{\perp}^2\left\{-\operatorname{Cov}[\beta(z),\gamma(z)]+\operatorname{Var}[\alpha(z)]\right\}, \label{uniformeps1} \\
\epsilon_{\rm int}^2&=\epsilon_{\perp}\left\{\operatorname{E}[\beta(z)]\operatorname{Var}[\mu_{p_x}(z)]+\operatorname{E}[\gamma(z)]\operatorname{Var}[\mu_x(z)]+2\operatorname{E}[\alpha(z)]\operatorname{Cov}[\mu_x(z),\mu_{p_x}(z)]\right\}. \label{uniformeps2}
\end{align}
\end{subequations}
Here we have used the facts that for any constant $a$ and any $z-$dependent functions $\phi$, $\phi_1$, and $\phi_2$:
\begin{equation}
\operatorname{E}[a\phi]=a\operatorname{E}[\phi],\quad \operatorname{Cov}[a\phi_1,a\phi_2]=a^2\operatorname{Cov}[\phi_1,\phi_2],\quad \operatorname{Var}[a\phi]=a^2\operatorname{Var}[\phi].
\end{equation}
Also, we used the fact that due to the assumption in (\ref{mean}):
\begin{equation}
\operatorname{Var}[\epsilon_x(z)]=\operatorname{E}[(\epsilon_x(z)-\operatorname{E}[\epsilon_x(z)])^2]=0.
\end{equation}
The expression (\ref{uniformeps1}) indicates that the mismatch emittance $\epsilon_R$ is determined by the size of slice-to-slice variations in the beam Twiss functions.  However, it is convenient (and somewhat surprising) that $\epsilon_R$ can also be written directly in terms of the {\it mean} Twiss functions.  This can be seen by noting from (\ref{Twiss1}) that:
\begin{align}
1&=\operatorname{E}[\beta(z)\gamma(z)-\alpha^2(z)] =\operatorname{E}[\beta(z)\gamma(z)]-\operatorname{E}[\alpha^2(z)]\\
&=\operatorname{Cov}[\beta(z),\gamma(z)]+\operatorname{E}[\beta(z)]\operatorname{E}[\gamma(z)]-\operatorname{Var}[\alpha(z)]-\operatorname{E}[\alpha(z)]^2,
\end{align}
so:
\begin{equation}
\epsilon_R^2=\epsilon_{\perp}^2\left\{\operatorname{E}[\beta(z)]\operatorname{E}[\gamma(z)]-\operatorname{E}[\alpha(z)]^2-1\right\}.
\end{equation}
When $\beta(z)$, $\gamma(z)$, and $\alpha(z)$ are each independent of $z$, we see that this quantity indeed vanishes as a consequence of (\ref{Twiss1}).

\subsection{Uniform slice Twiss functions}
Suppose that the Twiss functions of every longitudinal beam slice are identical.  More precisely, assume that $\epsilon_x(z)\neq 0$ and:
\begin{equation}
\alpha(z)=\operatorname{E}[\alpha(z)],\quad\beta(z)=\operatorname{E}[\beta(z)],\quad \gamma(z)=\operatorname{E}[\gamma(z)], \label{uniformTwiss}
\end{equation}
for every $z$ with $\lambda(z)\neq 0$.  We let $\alpha_0$, $\beta_0$, $\gamma_0$ denote the uniform values given in (\ref{uniformTwiss}).  The mismatch emittance in (\ref{decomp}) then becomes:
\begin{align}
\epsilon_R^2&=\operatorname{Var}[\epsilon_x(z)]-\operatorname{Cov}[\epsilon_x(z)\beta_0,\epsilon_x(z)\gamma_0]+\operatorname{Var}[-\epsilon_x(z)\alpha_0] \notag \\
&=\operatorname{Var}[\epsilon_x(z)]-\beta_0\gamma_0\operatorname{Cov}[\epsilon_x(z),\epsilon_x(z)]+\alpha_0^2\operatorname{Var}[\epsilon_x(z)] \notag \\
&=\operatorname{Var}[\epsilon_x(z)]-(\beta_0\gamma_0-\alpha_0^2)\operatorname{Var}[\epsilon_x(z)]=0,
\end{align}
since $\beta_0\gamma_0-\alpha_0^2=1$.  Thus, the misalignment emittance vanishes.  The contributions $\epsilon_{\perp}$ and $\epsilon_{||}$ are independent of the slice second moments, and their form remains unchanged.  The remaining contribution becomes:
\begin{align}
\epsilon_{\rm int}^2
&=\operatorname{E}[\epsilon_x(z)\beta_0]\operatorname{Var}[\mu_{p_x}(z)]+\operatorname{E}[\epsilon_x(z)\gamma_0]\operatorname{Var}[\mu_x(z)]-2\operatorname{E}[-\epsilon_x(z)\alpha_0]
\operatorname{Cov}[\mu_x(z),\mu_{p_x}(z)] \notag \\
&=\epsilon_{\perp}\left\{\beta_0\operatorname{Var}[\mu_{p_x}(z)]+\gamma_0\operatorname{Var}[\mu_x(z)]+2\alpha_0\operatorname{Cov}[\mu_x(z),\mu_{p_x}(z)]\right\}.
\end{align}

\subsection{Aligned slice centroids}
Next, suppose that all slice centroids are aligned.  More precisely, assume that:
\begin{equation}
\mu_x(z)=\operatorname{E}[\mu_x(z)],\quad\quad \mu_{p_x}(z)=\operatorname{E}[\mu_{p_x}(z)],
\end{equation}
for every $z$ with $\lambda(z)\neq 0$.  It immediately follows that:
\begin{equation}
\operatorname{Var}[\mu_x(z)]=\operatorname{Var}[\mu_{p_x}(z)]=\operatorname{Cov}[\mu_x(z),\mu_{p_x}(z)]=0, \label{aligned}
\end{equation}
and so $\epsilon_{\rm int}=\epsilon_{||}=0$.  Thus, the only projected emittance contributions are due to the mean slice emittance and the mismatch emittance.

 This case is often treated by introducing a {\it slice mismatch parameter}  \cite{Mismatch1}-\cite{Mismatch3}, given by:
\begin{equation}
\zeta(z)=\frac{1}{2}\left\{\gamma(z)\beta-2\alpha(z)\alpha+\beta(z)\gamma\right\}, \label{mismatch}
\end{equation}
where $\alpha$, $\beta$, and $\gamma$ are the projected Twiss functions, defined by:
\begin{equation}
\Sigma^X=\epsilon_x
\begin{pmatrix}
\beta & -\alpha \\
-\alpha & \gamma
\end{pmatrix}\quad\text{with}\quad \beta\gamma-\alpha^2=1,
\end{equation}
and $\epsilon_x$ is the projected emittance.  An alternative expression is given in terms of the projected and slice Twiss matrices, given respectively by:
\begin{equation}
R=\begin{pmatrix}
\beta & -\alpha \\
-\alpha & \gamma
\end{pmatrix}\quad\quad
R(z)=\begin{pmatrix}
\beta(z) & -\alpha(z) \\
-\alpha(z) & \gamma(z)
\end{pmatrix},
\end{equation}
from which (\ref{mismatch}) takes the form:
\begin{equation}
\zeta(z)=\frac{1}{2}\operatorname{tr}(R^{-1}R(z)).  \label{mismatch2}
\end{equation}
The slice mismatch parameter has the properties that $\zeta(z)\geq 1$, and $\zeta(z)=1$ if and only if $R(z)=R$.  (A proof is included in Appendix C.)  As a result, $\zeta(z)$ is a measure of how far the Twiss functions of slice $z$ vary from the projected Twiss functions of the beam.  

Multiplying (\ref{mismatch}) by the slice emittance at $z$ reveals that:
\begin{align}
\epsilon_x(z)\zeta(z)=\frac{1}{2}\left\{\Sigma_{22}(z)\beta+2\Sigma_{12}(z)\alpha+\Sigma_{11}(z)\gamma\right\}. \label{interm2}
\end{align}
Applying (\ref{aligned}) to (\ref{covar}), we see that in the present case the projected and slice covariance matrices are related by:
\begin{equation}
\Sigma_{ab}=\operatorname{E}[\Sigma_{ab}(z)],\quad\quad a,b=1,\ldots,6. \label{interm}
\end{equation}
Taking the average of (\ref{interm2}) across slices using (\ref{interm}) gives immediately the projected emittance:
\begin{align}
\operatorname{E}[\epsilon_x(z)\zeta(z)]=\frac{1}{2}\left\{\Sigma_{22}\beta+2\Sigma_{12}\alpha+\Sigma_{11}\gamma\right\}
=\epsilon_x\left\{\beta\gamma-\alpha^2\right\}=\epsilon_x.
\end{align}
It therefore follows that:
\begin{equation}
\epsilon_R^2=\operatorname{E}[\epsilon_x(z)\zeta(z)]^2-\operatorname{E}[\epsilon_x(z)]^2. \label{mismatchsimple}
\end{equation}
It is clear from (\ref{mismatchsimple}) that when $\zeta(z)=1$ for every slice (the perfectly matched case), then $\epsilon_R=0$.

\subsection{Linear slice centroid misalignment}
For the final case, assume a purely linear misalignment of the beam centroids.  More precisely, assume that:
\begin{equation}
\mu_x(z)=\mu_x(z_c)+(z-z_c)\mu'_x,\quad\quad \mu_{p_x}(z)=\mu_{p_x}(z_c)+(z-z_c)\mu'_{p_x}, \label{linear}
\end{equation}
for slices with $\lambda(z)\neq 0$.  Here $z_c=\operatorname{E}[z]$ denotes the longitudinal location of the beam centroid, and $\mu'_x$, $\mu'_{p_x}$ are constants.  
This is a case that occurs regularly in linear optics systems with nonzero transverse-longitudinal coupling.  

The quantities $\epsilon_{\perp}$ and $\epsilon_R$ are independent of the slice centroid values, and their form will therefore remain unchanged.  To evaluate the remaining two emittance contributions, note from (\ref{linear}) that:
\begin{equation}
\operatorname{Var}[\mu_x(z)]=\operatorname{Var}[z](\mu'_x)^2,\quad\quad \operatorname{Var}[\mu_{p_x}(z)]=\operatorname{Var}[z](\mu'_{p_x})^2,\quad \operatorname{Cov}[\mu_x(z),\mu_{p_x}(z)]=\operatorname{Var}[z]\mu'_x\mu'_{p_x}. \label{varlin}
\end{equation} 
An immediate consequence is that the nonlinear misalignment emittance vanishes:
\begin{equation}
\epsilon_{||}^2=\operatorname{Var}[\mu_x(z)]\operatorname{Var}[\mu_{p_x}(z)]-\operatorname{Cov}[\mu_x(z),\mu_{p_x}(z)]^2=0.
\end{equation}
Evaluating the linear misalignment emittance gives:
\begin{equation}
\epsilon_{\rm int}^2=\sigma_z^2\left\{\operatorname{E}[\sigma_x^2(z)](\mu'_{p_x})^2+\operatorname{E}[\sigma_{p_x}^2(z)](\mu'_x)^2-2\operatorname{E}[\langle{xp_x\rangle}_z] \label{result_lin}
\mu'_x\mu'_{p_x}\right\}. 
\end{equation}

To interpret (\ref{result_lin}), it is helpful to re-express the quantities involved using the projected beam moments.  To determine $\mu'_x$ and $\mu'_{p_x}$, for example, take
the covariance of (\ref{linear}) with the slice coordinate $z$ to see that:
\begin{align}
\operatorname{Cov}[z,\mu_x(z)]&=\operatorname{Cov}[z,\mu_x(z_c)]+\operatorname{Cov}[z,z-z_c]\mu'_x=\operatorname{Var}[z]\mu'_x,
\end{align}
with a corresponding expression involving $\mu'_{p_x}$.
However, it also follows from (\ref{covar}) that:
\begin{align}
\Sigma_{55}&=\operatorname{E}[\Sigma_{55}(z)]+\operatorname{Cov}[\mu_5(z),\mu_5(z)]=\operatorname{Var}[z], \\
\Sigma_{51}&=\operatorname{E}[\Sigma_{51}(z)]+\operatorname{Cov}[\mu_5(z),\mu_1(z)]=\operatorname{Cov}[z,\mu_x(z)],
\end{align} 
since for each slice $z$
\begin{equation}
\Sigma_{55}(z)=0,\quad\quad \Sigma_{51}(z)=0,\quad\quad \mu_5(z)=z.
\end{equation}
Thus we have:
\begin{equation}
\mu'_x=\frac{\Sigma_{51}}{\Sigma_{55}}, \quad\quad\mu'_{p_x}=\frac{\Sigma_{52}}{\Sigma_{55}}. \label{derivs}
\end{equation}
These quantities jointly characterize the size of transverse-longitudinal correlations in the beam.
Likewise, applying (\ref{covar}) together with (\ref{varlin}) gives:
\begin{subequations} \label{lincases}
\begin{align}
\Sigma_{11}&=\operatorname{E}[\Sigma_{11}(z)]+\operatorname{Cov}[\mu_1(z),\mu_1(z)]=\operatorname{E}[\sigma_x^2(z)]+\operatorname{Var}[z](\mu'_x)^2, \\
\Sigma_{22}&=\operatorname{E}[\Sigma_{22}(z)]+\operatorname{Cov}[\mu_2(z),\mu_2(z)]=\operatorname{E}[\sigma_{p_x}^2(z)]+\operatorname{Var}[z](\mu'_{p_x})^2, \\
\Sigma_{12}&=\operatorname{E}[\Sigma_{12}(z)]+\operatorname{Cov}[\mu_1(z),\mu_2(z)]=\operatorname{E}[\langle{\Delta x\Delta p_x\rangle}_z]+\operatorname{Var}[z]\mu'_x\mu'_{p_x}.
\end{align}
\end{subequations}
Using (\ref{lincases}) and (\ref{derivs}) to re-express (\ref{result_lin}) in terms of the projected beam covariance matrix gives:
\begin{equation}
\epsilon_{\rm int}^2=\frac{1}{\Sigma_{55}}\left(\Sigma_{11}\Sigma_{52}^2+\Sigma_{22}\Sigma_{51}^2-2\Sigma_{12}\Sigma_{51}\Sigma_{52}\right).
\end{equation}
Note that this quantity is defined only when $\Sigma_{55}=\sigma_z^2\neq 0$.  This is reasonable, since when $\sigma_z=0$, the decomposition of the beam into longitudinal slices is no longer meaningful.

\section{Generalizations and Related Results}
The decomposition given in (\ref{decomp}) can be generalized in a number of ways.  First, note that the treatment leading up to (\ref{decomp}) is independent of the phase space variables used, and the same result may be applied if the conjugate canonical variables $(z,p_z)$ are replaced by the conjugate variables $(t,p_t)$ or $(-ct,\delta)$, as are typically used to describe the beam distribution at a fixed longitudinal location, where for example:
\begin{equation}
p_t=-\gamma mc^2,\quad\quad \delta=\Delta\gamma/\gamma_0. \label{relativistic}
\end{equation}
In this case, the longitudinal beam slices are defined by the temporal coordinate $t$ (or $-ct$) rather than by the spatial coordinate $z$.  Alternatively, one may decompose the beam into slices based on the horizontal coordinate $x$ or the vertical coordinate $y$.  The following two cases are worth describing in some detail.

 \subsection{Emittance decomposition using 3-D spatial coordinates}
The discussion of Section 2 can be modified by replacing the decomposition of the beam into longitudinal slices (using the longitudinal coordinate $z$) with a decomposition of the beam into spatial points (using the three spatial coordinates $x$, $y$, and $z$).  
To see how this is done, we write ${\bf X}=({\bf r},{\bf p})$ to emphasize the separation of the spatial ${\bf r}=(x,y,z)$ and momentum  ${\bf p}=(p_x,p_y,p_z)$ phase space coordinates.

As in Section 2, the beam distribution function $f$ defines a probability density on the 6-D phase space, and it is natural to define both a marginal and a conditional probability density associated with the decomposition into spatial points.  In place of the marginal probability density given in (\ref{longdens}), we now have the 3-D spatial density:
\begin{equation}
\rho({\bf r})=\int f({\bf r},{\bf p})d{\bf p}.\label{spatialdensity}
\end{equation}
Likewise, in place of the conditional probability density given in (\ref{slicedist}), we now have the distribution function at a point ${\bf r}$:
\begin{equation}
f({\bf p}|{\bf r})=\frac{f({\bf r},{\bf p})}{\rho({\bf r})}, \label{spatialcondition}
\end{equation}
provided $\rho({\bf r})>0$, with $f({\bf p}|{\bf r})=0$ otherwise.  The averages $\langle{\cdot\rangle}_z$ and $\operatorname{E}[\cdot]$ are replaced by their analogues defined using densities (\ref{spatialdensity}) and (\ref{spatialcondition}), so we have the average of a quantity $\phi$ at a point ${\bf r}$ given by:
\begin{equation}
\langle{\phi\rangle}_{\bf r}=\int \phi({\bf r},{\bf p})f({\bf p}|{\bf r})d{\bf p}, \label{momavg}
\end{equation}
and the spatial average of any ${\bf r}-$dependent quantity $\phi$ given by:
\begin{equation}
\operatorname{E}[\phi]=\int \phi({\bf r})\rho({\bf r})d{\bf r}. \label{spatavg}
\end{equation}
In the derivation leading up to the emittance decomposition (\ref{decomp}), the conditional expectation and conditional covariance of Appendix A are now evaluated for a fixed triple of spatial coordinates ${\bf r}$, rather than for a fixed slice coordinate $z$.  Using the same theorems, the results of Section 3 carry over directly when one takes $z\rightarrow {\bf r}$ and replaces the averages $\langle{\cdot\rangle}_z$ and $\operatorname{E}[\cdot]$ with (\ref{momavg}) and (\ref{spatavg}).
 Note that in this case we have:
 \begin{equation}
 \mu_x({\bf r})=x,
 \end{equation}
 and the covariance matrix at point ${\bf r}$ takes the form:
 \begin{equation}
 \Sigma^X({\bf r})=\begin{pmatrix}
 0 & 0 \\
 0 & \sigma_{p_x}^2({\bf r})
 \end{pmatrix}. \label{spatialsigma}
 \end{equation}
 Of the four contributions to the projected emittance given in (\ref{decomp}), it follows from (\ref{spatialsigma}) that two of these always vanish:
 \begin{equation}
  \epsilon_{\perp}^2=\epsilon_R^2=0,
 \end{equation}
 and the remaining two take the forms:
 \begin{subequations}
 \begin{align}
 \epsilon_{\rm int}^2&=\operatorname{E}[\sigma_{p_x}^2({\bf r})]\operatorname{Var}[\mu_x({\bf r})], \label{intrinsic} \\
 \epsilon_{||}^2&=\operatorname{Var}[\mu_x({\bf r})]\operatorname{Var}[\mu_{p_x}({\bf r})]-\operatorname{Cov}[\mu_x({\bf r}),\mu_{p_x}({\bf r})]^2. \label{correl}
 \end{align}
 \end{subequations}
 The contribution (\ref{intrinsic}) is the {\it intrinsic} beam emittance due to intrinsic variations in the particle momentum at each point, while (\ref{correl}) is the {\it correlated} beam emittance due to spatially correlated variations in the particle momentum.  
 
 It is convenient to define the transverse beam temperature $T_x$ \cite{Reiser} at a point ${\bf r}$ by using the momentum spread at ${\bf r}$:
 \begin{equation}
 \frac{k_BT_x({\bf r})}{mc^2}=\sigma_{p_x}^2({\bf r}),
 \end{equation}
 where $k_B$ is Boltzmann's constant.  In this way, we see that (\ref{intrinsic}) is proportional to the spatial average of the transverse beam temperature times the square of the transverse rms beam size.  Meanwhile, (\ref{correl}) is the projected emittance that the beam would have {\it if it were cold}---that is, if all particles at a given point were given a momentum equal to the mean momentum $\mu_{p_x}({\bf r})$ at that point.

\subsection{Emittance decomposition using finite subpopulations}
In the previous sections, we described the beam by a continuous density $f$ on the phase space:  This is necessary for a rigorous discussion of slice moments and emittances, since these quantities are in general undefined when one considers a finite (discrete) particle distribution together with longitudinal slices of zero width.  However, the decomposition (\ref{decomp}) is based on mathematical tools that are quite general, which produce corresponding results for the discrete-particle case.  These results characterize the relationship between the rms emittances of the beam and the moments of its various subpopulations.

For example, suppose that a finite beam of $N$ particles is divided into $M$ disjoint bins $n=1,\ldots,M$.  These may be longitudinal bins of fixed or varying width, or bins based on some other criterion (such as particle energy); we assume only that the bins divide the phase space into non-overlapping regions.  We set $n=0$ for all points ${\bf X}$ in the phase space that lie outside the $M$ bins, so that each phase space point corresponds to a unique index $n$.  Then the value of this index defines a discrete random variable on the phase space, which plays exactly the role previously played by the continuous random variable $z$.  The conditional expectation and conditional covariance of Appendix A are now evaluated for fixed bin index $n$, rather than for fixed slice $z$.  Using the same theorems, the results of Section 3 carry over directly when one takes $z\rightarrow n$ and replaces the continuous averages $\langle{\cdot\rangle}$, $\langle{\cdot\rangle}_z$, and $\operatorname{E}[\cdot]$ with their discrete analogues.

In particular, let ${\bf X}_j$, $j=1,\ldots,N$ denote the particle phase space coordinates, and let $N_n$ denote the number of particles in bin $n$, with
\begin{equation}
\sum_{n=1}^{M}N_n=N.
\end{equation}
The averages in (\ref{avg}) and (\ref{sliceavg}) now become:
\begin{equation}
\langle{\phi\rangle}=\frac{1}{N}\sum_{j=1}^N\phi({\bf X}_j),\quad\quad \langle{\phi\rangle}_n=\frac{1}{N_n}\sum_{j=1}^{N_n}\phi({\bf X}_{j_n}),\quad\quad n=1,\ldots,M
\end{equation}
where the second sum is taken only over those particles that lie within bin $n$.  Thus, $\langle{\phi\rangle}_n$ denotes the bin average of the quantity $\phi$.  We may therefore define the bin centroid vectors $\vec{\mu}_n$, the bin covariance matrices $\Sigma_n$, and the bin emittances, in the natural way.
Finally, if $\phi_n$ is a quantity that varies from bin to bin, then we may define its average across bins as:
\begin{equation}
\operatorname{E}[\phi_n]=\frac{N_n}{N}\sum_{j=1}^M\phi_n,
\end{equation}
from which we may also define the covariance of two $n-$dependent quantities as in (\ref{zCovar}).  The horizontal projected emittance of the beam now takes the form:
\begin{equation}
\epsilon_x^2=\epsilon_{\perp}^2+\epsilon_R^2+\epsilon_{\rm int}^2+\epsilon_{||}^2,
\end{equation}
 where the four emittance contributions are given by (\ref{decomp}) after taking $z\rightarrow n$.
  
 \section{Conclusions}
The decomposition of a beam into longitudinal slices leads naturally to a corresponding decomposition for each of the projected emittances $\epsilon_x$, $\epsilon_y$, and $\epsilon_z$ into distinct nonnegative contributions, separating the effect of particle coordinate/momentum variation {\it within} each slice from the effect of particle coordinate/momentum variation {\it between} slices.   It is expected that each of these contributions is affected by distinct dynamical processes in a typical accelerator lattice, including  nonlinear aberrations due to external focusing fields, transverse-longitudinal coupling due to dispersion, slice mismatch due to time-dependent focusing effects, wakefield-induced emittance growth, and a variety of other effects.  As a result, independent consideration of these four contributions may provide helpful diagnostic tools for studying the dynamical sources of projected emittance growth in typical accelerator systems.  

\section{Acknowledgements}
This work was supported by the Director, Office of Science of the U.S. Department of Energy under Contract No. DE-AC02-05CH11231.

\appendix
\section{Appendix: Mathematical Theorems}
In this section, we describe several theorems that are useful in studying the relationships between slice and projected quantities.
\subsection{Theorems relating to probability}
The following is a fundamental theorem regarding conditional probability \cite{Probability1}-\cite{Probability2}.  We give the general result, and then prove the statement that is used in the text (\ref{expect}-\ref{covar}).

Suppose $X$, $Y$ and $Z$ are three random variables defined on the same probability space.  We use standard notation for the expectation $\operatorname{E}[\cdot]$ and covariance $\operatorname{Cov}[\cdot,\cdot]$ of these random variables.  The expectation of $X$ and the covariance of $X$ and $Y$ can be evaluated {\it under the condition that $Z$ takes on some fixed value $Z=z$}. The resulting quantities are written as:
\begin{equation}
\operatorname{E}[X|Z=z],\quad\quad \operatorname{Cov}[X,Y|Z=z],
\end{equation}
denoting the conditional expectation and the conditional covariance, respectively.
As the value of $Z$ varies, the functions $\operatorname{E}[X|Z]$ and $\operatorname{Cov}[X,Y|Z]$ are themselves random variables that depend on the value of $Z$.  The relationships between these random variables are given by the {\it law of total expectation}:
\begin{equation}
\operatorname{E}[X]=\operatorname{E}[\operatorname{E}[X|Z]]  \label{expect_gen}
\end{equation}
and the {\it law of total covariance}:
\begin{equation}
\operatorname{Cov}[X,Y]=\operatorname{E}[\operatorname{Cov}[X,Y|Z]]+\operatorname{Cov}[\operatorname{E}[X|Z],\operatorname{E}[Y|Z]], \label{covar_gen}
\end{equation}
which hold provided $\operatorname{E}[X]$ and $\operatorname{Cov}[X,Y]$ are each finite, respectively.
These results apply independently of whether the random variables $X$, $Y$, and $Z$ are continuous, discrete, or a mixture of the two.  

Consider now the phase space coordinates ${\bf X}$, with joint probability density given by the beam distribution function $f({\bf X})$.  Let $Z$ be the random variable corresponding to the coordinate $z$.  Then (\ref{expect_gen}) states that, for any function $\phi$ on the phase space:
\begin{equation}
\langle{\phi\rangle}=\operatorname{E}[\langle{\phi\rangle}_z]. \label{expect_notation}
\end{equation}
In the special case that $\phi({\bf X})=X_a$, this corresponds to the result quoted in the main text (\ref{expect}).  We now prove (\ref{expect_notation}).

Pf / Taking the mean of $\langle{\phi\rangle}_z$ across slices gives:
\begin{align}
\operatorname{E}[\langle{\phi\rangle}_z]&=\int \lambda(z)\langle{\phi\rangle}_zdz=\int \lambda(z)\left[\int  \phi(z,\zeta) f(\zeta|z)d\zeta\right]dz \notag \\
&=\int \left[ \int \phi(z,\zeta)f(z,\zeta)d\zeta\right]dz=\int \phi({\bf X}) f({\bf X})d{\bf X}=\langle{\phi\rangle},
\end{align}
where in the third equality we have made use of (\ref{slicedist}).  $\square$

Next, for some indices $a$ and $b$, let $X$ be the random variable corresponding to the coordinate $X_a$, let $Y$ be the random variable corresponding to the coordinate $X_b$, and let $Z$ be the random variable corresponding to the coordinate $z$.   Then we have, using the notation of the main text:
\begin{equation}
\operatorname{Cov}[X,Y|Z]=\Sigma_{ab}(z),\quad \operatorname{E}[X|Z]=\mu_a(z),\quad \operatorname{E}[Y|Z]=\mu_b(z),
\end{equation}
and (\ref{covar_gen}) corresponds to the result given in the main text as (\ref{covar}), a result that we now prove.

Pf / We have from the definition of the beam covariance matrix $\Sigma$ and (\ref{expect_notation}) that:
\begin{align}
\Sigma_{ab}&=\langle{X_aX_b\rangle}-\langle{X_a\rangle}\langle{X_b\rangle}=\operatorname{E}[\langle{X_aX_b\rangle}_z]-\operatorname{E}[\langle{X_a\rangle}_z]\operatorname{E}[\langle{X_b\rangle}_z]. \\
\intertext{Applying the definition of the slice centroid coordinates $\mu_a(z)$ and $\mu_b(z)$ gives:}
&=\operatorname{E}[\langle{(X_a-\mu_a(z))(X_b-\mu_b(z))\rangle}_z+\mu_a(z)\mu_b(z)]-\operatorname{E}[\mu_a(z)]\operatorname{E}[\mu_b(z)]. \\
\intertext{Applying the linearity of the expectation value and the definition of the slice covariance matrix:}
&=\operatorname{E}[\Sigma_{ab}(z)]+\operatorname{E}[\mu_a(z)\mu_b(z)]-\operatorname{E}[\mu_a(z)]\operatorname{E}[\mu_b(z)], \\
\intertext{which gives using (\ref{zCovar}) that:}
\Sigma_{ab}&=\operatorname{E}[\Sigma_{ab}(z)]+\operatorname{Cov}[\mu_a(z),\mu_b(z)].  \quad\quad\square
\end{align}

The following theorem makes use of the notion of convexity.  A {\it convex subset} of $n-$dimensional space $\mathbb{R}^n$ is a subset $C$ with the property that, for all ${\bf x}$ and ${\bf y}$ in $C$ and all $0\leq t\leq 1$, the point $t{\bf x}+(1-t){\bf y}\in C$.  (That is, the line segment joining any two points in $C$ is itself contained in $C$.)

If $C$ is a convex subset of $\mathbb{R}^n$, a {\it convex function on $C$} is a function $f:C\rightarrow \mathbb{R}$ such that, for all ${\bf x}$ and ${\bf y}$ in $C$ and all $0\leq t\leq 1$, 
\begin{equation}
f(t{\bf x}+(1-t){\bf y})\leq tf({\bf x})+(1-t)f({\bf y}). \label{convexity}
\end{equation}
Given a convex function $f:C\rightarrow \mathbb{R}$, a convex subset $K\subseteq C$ will be called an {\it $f$-affine subset} if and only if, for all ${\bf x}$ and ${\bf y}$ in $K$ and all $0\leq t\leq 1$, equality holds in (\ref{convexity}). 

{\it Jensen's inequality}:  Let $f$ be a convex function defined on a convex subset $C$ of $\mathbb{R}^n$, and let ${\bf X}$ be a random variable taking values in $C$.  Suppose that the mean $\operatorname{E}[{\bf X}]$ exists, in the sense that $\operatorname{E}[|X_j|]<\infty$ for each element $j=1,\ldots,n$.  It then follows that $E[{\bf X}]\in C$, the real number $\operatorname{E}[f({\bf X})]$ exists, and
\begin{equation}
f(\operatorname{E}[{\bf X}])\leq \operatorname{E}[f({\bf X})]. \label{Jensen}
\end{equation}
Furthermore, equality holds in (\ref{Jensen}) if and only if there exists an $f$-affine subset $K\subseteq C$ with the property that ${\bf X}\in K$ with probability 1 (a.e.).  The proof of (\ref{Jensen}) can be found in \cite{Jensen1}-\cite{Jensen2}.  The condition for equality follows from results described in \cite{Jensen3}-\cite{Jensen4}.

\subsection{Theorems relating to matrices}
A real $n\times n$ symmetric matrix $A$ is said to be {\it positive semidefinite} if ${\bf x}^TA{\bf x}\geq 0$ for all vectors ${\bf x}\in\mathbb{R}^n$, which is typically denoted by writing $A\geq 0$.
The following results are standard, and the reader may refer to one of the references \cite{Matrices1}-\cite{Matrices4} for proofs.

{\it Property 1}:  If $A$ is a positive semidefinite matrix, then the eigenvalues of $A$ are nonnegative and therefore $\operatorname{det}(A)\geq 0$.  If $\operatorname{det}(A)\neq 0$, then $A^{-1}$ is also positive semidefinite.

{\it Property 2}:  If $A$ is the covariance matrix associated with a vector of random variables, then $A$ is a positive semidefinite matrix.

{\it Property 3}:  The set of $n\times n$ positive semidefinite matrices is a convex subset of the space of real $n\times n$ matrices.

Pf / Let $A$ and $B$ be $n\times n$ positive semidefinite matrices, and let ${\bf x}\in\mathbb{R}^n$, so that:
\begin{equation}
{\bf x}^TA{\bf x}\geq 0,\quad\quad {\bf x}^TB{\bf x}\geq 0. \label{pdm}
\end{equation}
For any $0\leq t\leq 1$, the quantity $tA+(1-t)B$ is a real $n\times n$ symmetric matrix and:
\begin{equation}
{\bf x}^T(tA+(1-t)B){\bf x}=t({\bf x}^TA{\bf x})+(1-t)({\bf x}^TB{\bf x})\geq 0,
\end{equation}
so that $tA+(1-t)B$ is positive semidefinite.  $\square$


The following inequalities can be found in \cite{Matrices4}.

{\it Trace inequality}:  If $A$ and $B$ are $n\times n$ positive semidefinite matrices, then:
\begin{equation}
n\operatorname{det}^{1/n}(AB)\leq \operatorname{tr}(AB)\leq \operatorname{tr}(A)\operatorname{tr}(B). \label{Trace}
\end{equation}
Equality always holds when $n=1$.  Suppose $n>1$ and $\operatorname{det}(A)\neq 0$.  Then equality holds on the left iff $B=cA^{-1}$ for some $c\geq 0$, and equality holds on the right iff $B=0$, the zero matrix.

{\it Minkowski determinant inequality}:  If $A$ and $B$ are $n\times n$ positive semidefinite matrices, then
\begin{equation}
\operatorname{det}^{1/n}(A+B)\geq \operatorname{det}^{1/n}A+\operatorname{det}^{1/n}B. \label{Minkowski}
\end{equation}
Suppose $\operatorname{det}(A)\neq 0$.  Then equality holds iff $B=cA$ for some $c\geq 0$.

The following theorem can be viewed as a generalization of (\ref{Minkowski}) to a (possibly infinite) ensemble of matrices.

{\it Jensen's determinant inequality}:  Let $R$ be a random variable that takes its values in the set of $n\times n$ positive semidefinite matrices.  Suppose the expectation value $\operatorname{E}[R]$ exists, in the sense that:
\begin{equation}
\operatorname{E}[|R_{ab}|]<\infty,\quad a,b=1,\ldots,n.
\end{equation}
Then the matrix $\operatorname{E}[R]$ is also positive semidefinite, and satisfies:
\begin{equation}
\operatorname{det}^{1/n}\operatorname{E}[R]\geq \operatorname{E}[\operatorname{det}^{1/n}R]. \label{Jensendet}
\end{equation}
Furthermore, if the random matrix $R$ has the property that $\operatorname{det}(R)\neq 0$, then equality holds in (\ref{Jensendet}) if and only if there exists a real-valued random variable $c$ and a constant $n\times n$ matrix $R_0$ such that $R=cR_0$ with probability 1. 

Pf / Let $g$ be the function on the set of positive semidefinite matrices given by:
\begin{equation}
g(A)=-\operatorname{det}^{1/n}(A)
\end{equation}
for each $A\geq 0$.  Given any $0\leq t\leq 1$, it follows from (\ref{Minkowski}) that for any matrices $A\geq 0$ and $B\geq 0$:
\begin{align}
g(t A+(1-t)B)&=-\operatorname{det}^{1/n}(t A+(1-t)B) \leq -\operatorname{det}^{1/n}(t A)-\operatorname{det}^{1/n}((1-t)B) \notag \\
&= -t\operatorname{det}^{1/n}(A)-(1-t)\operatorname{det}^{1/n}(B)=t g(A)+(1-t)g(B).
\end{align}
Thus, the function $g$ is convex on the set of positive semidefinite matrices.  It now follows from Jensen's inequality (\ref{Jensen}) that $\operatorname{E}[R]\geq 0$ and:
\begin{equation}
g(\operatorname{E}[R])\leq \operatorname{E}[g(R)],
\end{equation}
from which (\ref{Jensendet}) immediately follows.  To prove the equality condition, it follows from (\ref{Minkowski}) that the only $g-$affine subsets in the set of positive semidefinite matrices with $\operatorname{det}(R)\neq 0$ are those of the form $\{cR_0: c\in [a,b]\}$ for some fixed matrix $R_0$.  Thus, equality holds in (\ref{Jensendet}) if and only if $R=cR_0$ with probability 1.  $\square$

\section{Appendix: Nonnegativity}
It follows from (\ref{slicemoment}) and (\ref{lambda}) that both $\Sigma^X(z)$ and $\Lambda^X$ take the form of covariance matrices.  By Property 2 of Appendix 6.2, each of these matrices is positive semidefinite, and we may apply all the theorems of that section.

1.  {\it Mean slice emittance} - Since $\Sigma^X(z)$ is positive semidefinite:
\begin{equation}
\epsilon_x(z)^2=\operatorname{det}\Sigma^X(z)\geq 0.
\end{equation}
Thus, $\epsilon_x(z)$ is a real quantity for each $z$, and so is the mean $\operatorname{E}[\epsilon_x(z)]$.  Then we must have:
\begin{equation}
\epsilon_{\perp}^2=\operatorname{E}[\epsilon_x(z)]^2\geq 0. \label{sliceinequality}
\end{equation}
Furthermore, since $\epsilon_x(z)\geq 0$ for all $z$, equality holds in (\ref{sliceinequality}) if and only if $\epsilon_x(z)=0$ a.e.

2.  {\it Mismatch emittance} - Recall from (\ref{eperp}) that:
\begin{equation}
\operatorname{det}(\operatorname{E}[\Sigma^X(z)])=\epsilon_{\perp}^2+\epsilon_R^2.
\end{equation}
The $2\times 2$ covariance matrix $\Sigma^X(z)$ is positive semidefinite for each $z$.  It follows from Jensen's determinant inequality (\ref{Jensendet}) that the matrix $\operatorname{E}[\Sigma^X(z)]$ is also positive semidefinite and satisfies:
\begin{equation}
\operatorname{det}^{1/2}(\operatorname{E}[\Sigma^X(z)])\geq \operatorname{E}[\operatorname{det}^{1/2}(\Sigma^X(z))]=\operatorname{E}[\epsilon_x(x)]=\epsilon_{\perp}.
\end{equation}
Thus,
\begin{equation}
\epsilon_R^2=\operatorname{det}(\operatorname{E}[\Sigma^X(z)])-\epsilon_{\perp}^2\geq 0. \label{mismatchinequality}
\end{equation}
Suppose that the slice emittance is nonzero, $\epsilon_x(z)>0$.  Then equality holds in (\ref{mismatchinequality}) if and only if $\Sigma^X(z)=c(z)R_0$ a.e. for some real-valued function $c(z)$ and a fixed matrix $R_0$.  We can assume without loss of generality that $\operatorname{det}(R_0)=1$ (otherwise we may scale $c$ appropriately).  Thus, $R_0=R(z)$ is exactly the slice Twiss matrix and $c(z)=\epsilon_x(z)$ is the slice emittance.  Hence, equality holds in (\ref{mismatchinequality}) if and only if the slice Twiss functions $\alpha(z)$, $\beta(z)$, and $\gamma(z)$ are constant a.e.

3.  {\it Linear misalignment emittance} - Let $A$ and $B$ be the $2\times 2$ matrices defined in (\ref{step1}):
\begin{equation}
A=\operatorname{E}[\Sigma^X(z)],\quad\quad B=\Lambda^X.
\end{equation}
Then $A$ and $B$ are each positive semidefinite.  By the trace inequality (\ref{Trace}), we have:
\begin{equation}
\epsilon_{\rm int}^2=\operatorname{tr}(A)\operatorname{tr}(B)-\operatorname{tr}(AB)\geq 0. \label{criterionlinear}
\end{equation}
Suppose that the slice emittance is nonzero, $\epsilon_{\perp}\neq 0$.  Then $\operatorname{det}(A)=\epsilon_{\perp}^2+\epsilon_{R}^2\neq 0$, and equality in (\ref{criterionlinear}) holds if and only if $B=\Lambda^X=0$, the zero matrix.  This implies that $\operatorname{Var}[\mu_x(z)]=\operatorname{Var}[\mu_{p_x}(z)]=0$, so that $\mu_x(z)=\operatorname{E}[\mu_x(z)]$ and $\mu_{p_x}(z)=\operatorname{E}[\mu_{p_x}(z)]$ a.e.  That is, the slice centroids are aligned.  In this case, we also have $\epsilon_{||}^2=\operatorname{det}(\Lambda^X)=0$.

4.  The matrix $\Lambda^X$ is positive semidefinite, and therefore:
\begin{equation}
\epsilon_{||}^2=\operatorname{det}(\Lambda^X)\geq 0. \label{nonlinearineq}
\end{equation}
Suppose that strict inequality holds in (\ref{nonlinearineq}), so that $\epsilon_{||}\neq 0$.  By the results of Section 5.4, it follows that the slice centroids $\mu_x(z)$, $\mu_{p_x}(z)$ are not uniform in $z$, nor are they described by linear functions of the form (\ref{linear}).  That is, one or both of $\mu_x(z)$, $\mu_{p_x}(z)$ possess a nonlinear dependence on the slice coordinate $z$.

\section{Appendix: Slice mismatch parameter}
Let $\zeta(z)$ be the slice mismatch parameter defined in (\ref{mismatch2}).  Then $\zeta(z)\geq 1$, and $\zeta(z)=1$ if and only if $R(z)=R$.

Pf / Since $\Sigma^X$ and $\Sigma^X(z)$ are positive semidefinite matrices and we have assumed that $\epsilon_x>0$, $\epsilon_x(z)>0$, the matrices $R=\Sigma^X(z)/\epsilon_x$ and $R(z)=\Sigma^X(z)/\epsilon_x(z)$ are also positive semidefinite.  By Property 1 of Appendix A, this is also true of $R^{-1}$.  Furthermore, we have:
\begin{equation}
\operatorname{det}(R)=\operatorname{det}(R(z))=1.
\end{equation}
It follows from the trace inequality (\ref{Trace}) that:
\begin{equation}
\zeta(z)=\frac{1}{2}\operatorname{tr}(R^{-1}R(z))\geq \operatorname{det}^{1/2}(R^{-1}R(z))=1. \label{mismatchineq}
\end{equation}
In addition, equality holds if and only if $R(z)=cR$ for some $c\geq 0$.  However, we also have:
\begin{equation}
1=\operatorname{det}(R(z))=\operatorname{det}(cR)=c^2\operatorname{det}(R)=c^2,
\end{equation}
so $c=\pm 1$.  Thus, $c=1$ and $R(z)=R$.  $\square$

\end{document}